\documentclass[prl,twocolumn,superscriptaddress,showpacs]{revtex4}
\usepackage{graphicx,amsmath,amssymb,bm}

\def\be{\begin{equation}}
\def\ee{\end{equation}}
\def\ba{\begin{eqnarray}}
\def\ea{\end{eqnarray}}
\def\bas{\begin{eqnarray*}}
\def\eas{\end{eqnarray*}}

\begin{document}

\title{Condensates of $p$-wave pairs are exact solutions for rotating two-component Bose gases}

\author{T.~Papenbrock}

\affiliation{Department of Physics and Astronomy, University of
  Tennessee, Knoxville, TN 37996, USA}

\affiliation{Physics Division, Oak Ridge National Laboratory, Oak
  Ridge, TN 37831, USA}

\affiliation {GSI Helmholtzzentrum f\"ur Schwerionenforschung GmbH,
  D-64291 Darmstadt, Germany}

\affiliation{Institut f\"ur Kernphysik, Technische Universit\"at
  Darmstadt, D-64289 Darmstadt, Germany}

\author{S. M. Reimann}

\affiliation{Mathematical Physics, LTH, Lund University, P.O.
  Box 118, SE-22100 Lund, Sweden}

\author{G. M. Kavoulakis}

\affiliation{Technological Educational Institute of Crete, P.O. Box
  1939, GR-71004, Heraklion, Greece}

\begin{abstract}
  We derive exact analytical results for the wave functions and
  energies of harmonically trapped two-component Bose-Einstein
  condensates with weakly repulsive interactions under rotation. The
  isospin symmetric wave functions are universal and do not depend on
  the matrix elements of the two-body interaction. The comparison with
  the results from numerical diagonalization shows that the ground
  state and low-lying excitations consists of condensates of $p$-wave
  pairs for repulsive contact interactions, Coulomb interactions, and
  the repulsive interactions between aligned dipoles.
\end{abstract}


\maketitle 

Exact analytical solutions for interacting quantum many-body systems
are very rare~\cite{Calogero,Dukelsky}. However, they are of
tremendous interest since they may provide us with further insight
into the correlations of quantum-mechanical many-body systems.  Even
rarer are cases where the exactly solvable quantum many-body system
can also potentially be realized experimentally.  A famous example is
the celebrated Laughlin state~\cite{Laughlin} and its
generalizations~\cite{Jain} that describe the two-dimensional electron
gas in the quantum Hall regime.  In recent years, the advances with
ultra-cold atomic quantum gases have significantly broadened the range
of experimentally accessible many-body systems that are also solvable
exactly.  An example is the interacting Bose gas in one dimension.
Here, the Bethe ansatz offers an analytic solution to the Lieb-Liniger
model~\cite{Lieb}, and the fermionization of bosons in the
Tonks-Girardeau~\cite{Girardeau} limit has been
observed~\cite{Paredes,Kinoshita}.

Exact analytical solutions exist also for single-component dilute and
weakly interacting Bose-Einstein condensates under
rotation~\cite{WGS,Mottelson,BP1999,JK,SmithWilkin,PB2001jpa,Huang,HV2,Reimann}.
For a large class of repulsive two-body interactions, the exact ground
state of $N$ bosons at angular momentum $L$ results from projecting
the unique state where $L$ particles carry one unit of angular
momentum onto the subspace with zero angular momentum for the center
of mass~\cite{PB2001jpa,Huang,HV2}.  In this Letter, we generalize
this exact solution to the interesting case of a two-component Bose
gas~\cite{Xie, Bargi2007}, and arrive at a very appealing result: As in the single-component
case, the exact solutions at angular momentum $L$ are states where $L$
bosons carry one unit of angular momentum each, and as before one has
to project out excitations of the center of mass.  In the
two-component case, however, isospin enters as a good quantum number, and 
eigenstates can be labeled by the number of isospin-singlet $p$-wave
pairs that enter the wave function. Isospin was introduced in nuclear
physics by Heisenberg~\cite{Heisenberg} and -- like ordinary spin --
is based on SU(2) symmetry of two-component systems. The comparison
with numerical results shows that for repulsive contact interactions,
Coulomb forces, and repulsive forces between aligned dipoles~\cite{Lahaye}, the
ground state contains a maximum number of isospin-singlet $p$-wave
pairs. This Letter also explains the recent study by Bargi {\it et
  al.}~\cite{Bargi2007}, who found by numerical diagonalization that
the yrast energy of two-component Bose gases is a simple function of
angular momentum.

We consider a harmonically trapped two-component dilute gas of $N$
bosons of a first species $\downarrow$, and $M$ bosons of a second
species $\uparrow$.  We assume that the interactions are
perturbatively weak and of equal strength between intra-species and
inter-species. This special case of equal scattering lengths is
approximately realized in gases of $^{87}$Rb and gases of
$^{23}$Na~\cite{Kasamatsu}.  Let us list the conserved quantities
involved in this problem. Besides the isospin $T$, the total angular
momentum $L$, the angular momentum of the center of mass $L_c$, the
total number of bosons $A\equiv M+N$ and the isospin projection
$T_z=(M-N)/2$ are conserved. There is no simple basis that reflects
these symmetries simultaneously. In second quantization the
conservation of $L$, $A$, $T_z$, and $T$ is straight forward, while
the conservation of $L$, $M$, $N$, and $L_c$ is most easily expressed
in the configuration space within first quantization.  We will employ
both pictures in what follows.

Single-particle states $\phi_{nlm}$ of the spherical harmonic
oscillator have energies $E_{nl}=\hbar\omega (2n+l+3/2)$. We are
interested in low-energetic states at high angular momentum. For
weakly perturbative interactions, only single-particle states with no
radial excitation ($n=0$) can contribute, and we can limit ourselves
to maximally aligned states with $m=l$. Thus, single-particle states
with single-particle angular momentum $l$ are
$\phi_{0ll}(z)=z^l\exp{(-|z|^2/2)}$ and
$\phi_{0ll}(w)=(w^l\exp{(-|w|^2/2)}$ for the bosons of the species
$\downarrow$ and $\uparrow$, respectively. We denote the coordinates
of bosons belonging to the $\downarrow$ species by $z_j=x_j+iy_j$
($j=1,\ldots,N$) and and by $w_j=u_j+iv_j$, ($j=1,\ldots,M$) for the
$\uparrow$ species. Here, $x$ and $y$ ($u$ and $v$) are the two
Cartesian coordinates perpendicular to the axis of rotation for the
$\downarrow$ ($\uparrow$) species. In what follows, we omit the
ubiquitous Gaussians from the single-particle wave functions. Let
$\hat{b}_{l\downarrow}^\dagger$ and $\hat{b}_{l\uparrow}^\dagger$
create a boson in the state corresponding to $z^l$ and $w^l$,
respectively. The corresponding annihilation operators are
$\hat{b}_{l\downarrow}$ and $\hat{b}_{l\uparrow}$, and the creation
and annihilation operators fulfill the canonical commutation relations
for bosons.

In the case of a single species of bosons under rotation, the ground
state at angular momentum $L$ consists essentially of $L$ bosons
carrying one unit of angular momentum each while the remaining $N-L$
bosons carry no angular momentum~\cite{JK}. Small modifications of
this picture are due to the preservation of the angular momentum of
the center of mass. Remarkably, the two-component case is somewhat
similar, and Bargi {\it et al.}~\cite{Bargi2007} numerically found
that the ground state consists entirely of single-particle states with
angular momenta $l=0$ and $l=1$.  For $L\le N\le M$, there are $L+1$
such states (labeled by the number of bosons of the species
$\downarrow$ that are in the single-particle state with $l=1$). What
is the ground state for repulsive interactions in the space spanned by
these states? To address this question for a two-component system, we
recall the essence of Hund's rule: For repulsive interactions, the
interaction energy is minimized for wave functions that are
antisymmetric in position space. In fermionic electron systems such as
atoms and quantum dots, this leads to a symmetric spin wave function.
The same reasoning applied to the present case of a two-species Bose
gas would require the isospin wave function to be antisymmetric, too,
thus making the total wave function symmetric under particle exchange.
The operator
\be
\hat{B}^\dagger \equiv {1\over\sqrt{2}}\left(\hat{b}^\dagger_{1\downarrow}
\hat{b}^\dagger_{0\uparrow} - \hat{b}^\dagger_{0\downarrow}\hat{b}^\dagger_{1\uparrow}\right)
\ee
creates an isospin-singlet $p$-wave pair (i.e. $L=1$ and $T=T_z=0$).
This pair is antisymmetric in position space and antisymmetric in
isospin space and thus totally symmetric under exchange. Thus, the
interaction energy is minimized for condensates of isospin-singlet
pairs.  In general, we have $N\ne M$, and the ground-state wave
function will also consist of unpaired bosons. The states
\be
\label{hi}
|\chi_\tau\rangle \equiv
\left(\hat{T}_-\right)^{N-\tau}\left(\hat{b}^\dagger_{0\uparrow}\right)^{A-L-\tau}
\left(\hat{b}^\dagger_{1\uparrow}\right)^{L-\tau}
\left(\hat{B}^\dagger\right)^\tau |0\rangle 
\ee 
have angular momentum $L$, isospin $T=A/2-\tau$, isospin projection
$T_z=(M-N)/2$, and consist entirely of single-particle states with
angular momenta $l=0$ and $l=1$. Here $|0\rangle$ denotes the vacuum,
and $\tau=0,1,\ldots\min{(L,N)}$ is the number of isospin-singlet
pairs.  The isospin operators are $\hat{T}_z=\sum_{l=0}^\infty
(\hat{b}^\dagger_{l\uparrow}\hat{b}_{l\uparrow}-\hat{b}^\dagger_{l\downarrow}\hat{b}_{l\downarrow})/2
$,
$\hat{T}_-=\sum_{l=0}^\infty\hat{b}^\dagger_{l\downarrow}\hat{b}_{l\uparrow}$,
and $\hat{T}^2 = \hat{T}_-\hat{T}_-^\dagger + T_z(T_z+1)$.  The
eigenvalues of $\hat{T}_z$ and $\hat{T}^2$ are denotes as $T_z$ and
$T(T+1)$, respectively. Let us understand the state~(\ref{hi}) in
detail starting from the right.  The application of the pair operators
$\hat{B}^\dagger$ to the vacuum yields a state of $2\tau$ bosons with
quantum numbers $L=\tau, T=0, T_z=0$. The operators
$\hat{b}_{1\uparrow}^\dagger$ yields the angular momentum $L$ we seek,
increase the number of bosons to $L+\tau$, while keeping isospin
$T=T_z=(L-\tau)/2$ a good quantum number.  The application of the
operators $\hat{b}_{0\uparrow}^\dagger$ increase the number of bosons
to $A$ and keeps isospin a good quantum number.  Finally, the desired
particle numbers $M$ and $N$ (i.e. $T_z=(M-N)/2)$ results from the
$\hat{T}_-$ operators. For repulsive interactions the ground state
consists of the maximum number of isospin-singlet pairs (i.e.
$\tau=\min{(L,N)}$). Thus, the observation by Bargi {\it et al.},
together with an adaptation of Hund's rule for bosons and isospin
symmetry leads to eigenstates~(\ref{hi}) consisting of condensates of
isospin-singlet $p$-wave pairs. Note that these arguments are
independent of the details of the repulsive interaction.  These are
the main result of the present Letter.  As in the single-component
case, minor modifications of this picture are due to the conserved
angular momentum of the center of mass. Let us contrast our results to
BCS pairing in Fermi systems. In BCS theory, a weakly attractive
interaction leads to the formation of Cooper pairs (i.e.  pairs of
fermions in time-reversed orbits). The resulting BCS ground state is a
condensate of spin-singlet $s$-wave pairs, where the symmetric
configuration-space wave function optimizes the interaction energy. In
our case, we deal with bosons and with repulsive interactions, and
this modifies the picture accordingly.

Let us compute the energies of the states~(\ref{hi}). We generalize and
extend the results of Ref.~\cite{PB2001jpa} to the case of
two-component Bose gases and sketch the main steps. The Hilbert space
${\cal H}_L^{(N)}$ of $N$ identical bosons $\downarrow$ at total
angular momentum $L$ is spanned by products
$e_{\lambda_1}(z)e_{\lambda_2}(z)\cdot\ldots\cdot e_{\lambda_k}(z)$
(with $\lambda_1+\lambda_2+\ldots+\lambda_k=L$) of elementary
symmetric polynomials
\be 
e_\lambda(z) \equiv
e_\lambda(z_1,\ldots,z_N) = \sum_{1\le i_1<...<i_\lambda\le N} z_{i_1}
z_{i_2}\cdots z_{i_\lambda} \ .  
\ee 
Note that $e_\lambda(z)$ carries
$\lambda$ units of angular momentum.  For two-component mixtures of
$N$ and $M$ identical bosons with $N\le M$, the Hilbert space at total
angular momentum $L$ (with $L\le M$) is the sum 
\be
\label{hilbert}
\sum_{\lambda=0}^{{\rm min}(L,N)} {\cal H}_\lambda^{(N)} \otimes {\cal
  H}_{L-\lambda}^{(M)} 
\ee 
of direct products of the Hilbert spaces of each component. Products
of elementary symmetric polynomials are linearly independent and form
a basis. In the absence of interactions, all states in the Hilbert
space are degenerate. Perturbatively weak interactions will lift this
degeneracy.

The two-body interaction for a two-component mixture of Bose gases can
be written as
\be
\label{ham}
V=\sum_{m\ge 0} v_m\hat{V}_m=\sum_{m\ge 0}v_m(\hat{A}_m + \hat{B}_m +\hat{C}_m) \ .
\ee
Here, $v_m$ is a matrix element and 
\ba
\label{abc}
\hat{A}_m &=& \sum_{1\le i<j\le N} (z_i-z_j)^m(\partial_{z_i}-\partial_{z_j})^m  \ , \\
\hat{B}_m &=& \sum_{1\le i<j\le M} (w_i-w_j)^m(\partial_{w_i}-\partial_{w_j})^m \ , \\
\hat{C}_m &=& \mathop{\sum_{1\le i\le N}}_{1\le j\le M} (z_i-w_j)^m(\partial_{z_i}-\partial_{w_j})^m  
\ea
are the interactions between bosons of species $\downarrow$, between
bosons of species $\uparrow$, and the inter-species interaction,
respectively. By construction, the interaction preserves angular
momentum $L$ (i.e. the degree of the monomial wave function it is
acting on), is of two-body nature, and -- due to its translationally
invariant form (only differences of coordinates and derivatives
appear) -- preserves the angular momentum $L_c$ of the center of mass.
Furthermore, the interaction~(\ref{ham}) is invariant under the
exchange of $z_l\leftrightarrow w_k$ and therefore preserves isospin.
For the zero-ranged contact interaction, we have
$v_m=(-1/2)^m/m!$~\cite{PB2001jpa}.

The action of the operators~(\ref{abc}) on elementary
symmetric polynomials is particularly simple~\cite{PB2001jpa}
\ba
\label{relations}
\hat{A}_m e_\lambda(z) = \hat{B}_m e_\mu(w) = 0 &&\quad\mbox{for $m\ge 3$} \ ,\nonumber\\
\hat{C}_m e_\lambda(z) e_\mu(w)=  0 &&\quad\mbox{for $m\ge 3$} \ ,\nonumber\\
\hat{A}_m e_1(z) = \hat{B}_m e_1(w) =\hat{C}_m R = 0 &&\quad\mbox{for $m\ge 1$} \ .
\ea
Here, $R\equiv {1\over A} \left(e_1(z)+e_1(w)\right)$ denotes the
center of mass.  Equations~(\ref{relations}) show that only the terms
$0\le m\le 2$ of the Hamiltonian~(\ref{ham}) are of interest when
acting on products $e_\lambda(z) e_\mu(w)$. For the operators $V_0$
and $V_1$ we find
\ba
\hat{V}_0&=&A(A-1)/2 \ , \quad \hat{V}_1 = A(\hat{L}-\hat{L}_c) \ , \nonumber\\
\hat{L}&\equiv&\sum_{i=1}^N z_i\partial_{z_i} +\sum_{j=1}^M w_j\partial_{w_j} \ , \nonumber\\ 
\hat{L}_c&\equiv&{1\over A}\left(\sum_{i,j=1}^N\sum_{k,l=1}^M  z_i w_k\partial_{z_j}\partial_{w_l}\right) \ . 
\ea
Only $\hat{V}_2$ is truly non-trivial, and
\ba
\label{key}
\lefteqn{\hat{V}_2 e_\lambda(z) e_{\mu}(w)=
2(\lambda N +\mu M +2\lambda\mu) e_\lambda(z) e_{\mu}(w)} \nonumber\\
&+&2(N-\lambda+1)(\mu+1) e_{\lambda-1}(z) e_{\mu+1}(w) \nonumber\\
&+&2(M-\mu+1) (\lambda+1) e_{\lambda+1}(z) e_{\mu-1}(w) \ , \nonumber\\
&-&2A(N-\lambda+1) e_{\lambda-1}(z) e_{\mu}(w) R \nonumber\\
&-&2A(M-\mu+1) e_{\lambda}(z) e_{\mu-1}(w) R\ .
\ea
Equation~(\ref{key}) shows that the set
\ba
\label{M}
{\cal M}\equiv \{R^{n}e_\lambda(z)e_{L-\lambda-n}(w)\} \quad\mbox{with}\\
0\le \lambda\le {\rm min}(L,N) \ , \quad 0\le n \le L-\lambda\nonumber
\ea
spans a subspace in Hilbert space at angular momentum $L$ that is left
invariant by $\hat{V}_2$.  This subspace contains the states
$e_\lambda(z) e_{L-\lambda}(z)$ which are linear combinations of the
eigenstates~(\ref{hi}). The states $e_\lambda(z) e_{L-\lambda}(z)$,
are, however, not eigenstates of the center-of-mass momentum.  Let
$\hat{P}_0$ be the projector onto wave functions with zero angular
momentum of the center of mass, i.e.
$\hat{P}_0\psi(z,w)=\psi(z-R,w-R)$ for any wave function $\psi(z,w)$.
The wave functions $P_0 e_\lambda(z) e_{L-\lambda}(z)$ are in the
subspace spanned by ${\cal M}$~\cite{PB2001jpa}.  Thus, the states
$\hat{P}_0|\chi_\tau\rangle$ with $|\chi_\tau\rangle$ with from
Eq.~(\ref{hi}) are eigenstates of the Hamiltonian~(\ref{ham}). Note
that these states do not depend on the matrix elements $v_m$ of the
interaction.

To compute the corresponding eigenvalues we make the ansatz
$P_0\psi_{L,n}$ for the eigenfunction with
\be
\label{solpsi_gen}
\psi_{L,n} = \sum_{\lambda=0}^{{\rm min}(L,N)} 
c^{(n)}_\lambda e_\lambda(z) e_{L-\lambda}(w) \ .
\ee
Here, $n$ is an additional label that distinguishes between ${\rm
  min}(L,N)+1$ different wave functions. Our results below suggest
that $n$ is the number of isospin-singlet pairs.  The eigenvalue
equation $V P_0\psi_{L,n} = E_n P_0\psi_{L,n}$ requires the
coefficients $c_\lambda^{(n)}$ to fulfill
\ba
\label{ev}
0&=&\left(LM +\lambda(2L+N-M-2\lambda)-{\varepsilon_n}\right)c_\lambda^{(n)} \\
&+&(N-\lambda)(L-\lambda)c_{\lambda+1}^{(n)} +\lambda(M-L+\lambda) c_{\lambda-1}^{(n)} 
\nonumber \ .
\ea 
Here, $\varepsilon$ enters the energy eigenvalue
\ba
\label{spec}
E_n=A(A-1){v_0\over 2} + A L v_1 +2v_2\varepsilon_n \ . 
\ea
Note that the eigenvalue problem~(\ref{ev}) does not depend on the
matrix elements of the two-body interaction. For the solution of the
eigenvalue problem, we make the ansatz
\be
\label{solc_gen}
c_\lambda^{(n)} = \sum_{k=0}^n \beta_k\lambda^k \ .  
\ee 
Here, we concealed the fact that the coefficients $\beta_k$ also
depend on $n$.  We insert the ansatz~(\ref{solc_gen}) into
Eq.~(\ref{ev}) and compare the coefficients of $\lambda^m$,
$m=0,\ldots n+2$. This yields 
\be 
\varepsilon_n = AL  -n(A+1-n) \ ,
\ee 
which enters the energy~(\ref{spec}). The coefficients $\beta_{k},
k<n$ are recursively defined in terms of $\beta_n$ (which sets the
normalization).  The quantum number $n$ acquires the values
$n=0,1,2,\ldots {\rm min}(L,N)$, and the lowest energy is obtained for
$n={\rm min}(L,N)$.  

\begin{figure}[h]
        \includegraphics[width=0.45\textwidth,angle=0]{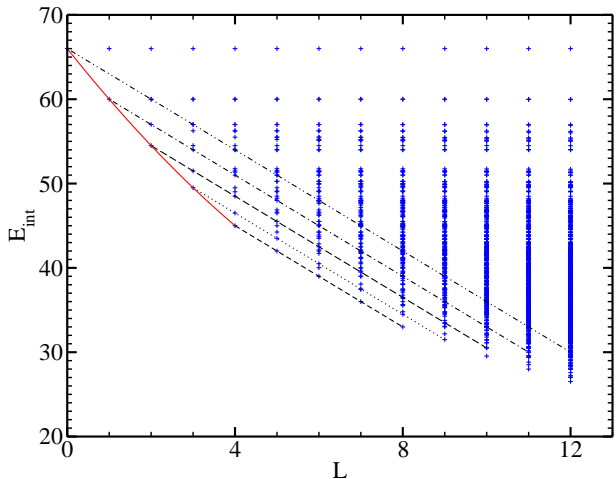}
        \caption{(Color online) Spectrum of a two-component Bose gas
          with $N=4$ and $M=8$ bosons per species, respectively, as a
          function of the angular momentum $L$ for the contact
          interaction. The broken lines connect states with energies
          $E_n$ from Eq.~(\ref{spec}) for fixed $n=0,1,2,\ldots N$
          (from top to bottom). The solid (red) line connects states
          with energies $E_L$ for $L\le N$.}
    \label{fig1}
\end{figure}

Figure~\ref{fig1} shows the energy spectrum of a two-component
Bose-Einstein condensate with $N=4$ and $M=8$ particles per species,
respectively, as a function of the angular momentum $L$. The broken
lines connect states with energies $E_n$ from Eq.~(\ref{spec}) for
fixed $n=0,1,2,\ldots$ (from top to bottom).  The ground state has
$n=\min{(L,N)}$. Note that $c_\lambda^{(0)}=1$ solves the eigenvalue
problem~(\ref{ev}) and yields the state
$\psi_{L,0}=e_L(z_1,\ldots,z_N,w_1,\ldots,w_M)$.  This state is
totally symmetric under the exchange of any particles, has maximum
isospin $T=A/2$ and contains no isospin-singlet $p$-wave pairs.  For
$L\le N$, the ground state~(\ref{solpsi_gen}) with $n=L$ has
coefficients $c_\lambda^{(L)}/c_0^{(L)} = (-1)^\lambda (M-L+\lambda)!
(N-\lambda)!/ [(M-L)! N!]$ and can be rewritten as $\hat{S}
\prod_{k=1}^L (z_{k}-w_{k})$.  Here the symmetrization operator $\hat{S}$ ensures
the symmetry under exchange of bosons of each species, and the
antisymmetry between the two species in position space is evident. We
thus believe that the label $n$ of the energies~(\ref{spec}) has to be
identified with the number $\tau$ of isospin-singlet $p$-wave pairs of the
eigenstates $\hat{P}_0 |\chi_\tau\rangle$ with $|\chi_\tau\rangle$
from Eq.~(\ref{hi}).

The reasoning that led to the isospin-singlet $p$-wave
condensates~(\ref{hi}) was based on general arguments regarding
repulsive interactions in SU(2)-symmetric two-component systems.  To
check our arguments, we performed numerical computations for Coulomb
interactions and repulsive interactions between aligned dipoles.  For
the Coulomb interaction, the relevant matrix elements are
$v_0=\sqrt{\pi/2}$, $v_1=-v_0/4$, and $v_2=3v_0/64$, respectively, and
we refer the reader to Ref.~\cite{HV2} for the analytical derivation.
The comparison with numerical results shows that the exact
results~(\ref{spec}) again describe the ground states and low-lying
excitations.  Finally, we consider repulsive interactions between
dipoles aligned perpendicular to the trap plane.  The comparison of
the numerical spectra and the analytical results~(\ref{spec}) yields
$v_0\approx6.868$, $v_1\approx-3.188$, and $v_2 = 0.755$,
respectively. Again, the ground state and low-lying excitations are
described by the analytical results.

In summary, we showed that low-lying states of rotating two-component
Bose gases with weak repulsive interactions are condensates of
isospin-singlet $p$-wave pairs, and we derived analytical expressions
for the energies and the corresponding wave functions.  The wave
functions are universal as they do not depend on the details of the
two-body interaction.  Numerical computations demonstrate that these
eigenstates are the ground states and some of the low-lying
excitations for the contact interaction, the Coulomb interaction, and
repulsive interactions between aligned dipoles.

\section*{Acknowledgments}
We thank S. Bargi, J. Cremon, and W. Nazarewicz for discussions. We
also thank J. Cremon for assistance with the numerical work.  This
work was partly supported by the U.S. Department of Energy under Grant
No. DE-FG02-96ER40963, by the Alexander von Humboldt-Stiftung, and by
the Swedish Research Council.

\end{document}